\documentclass[lettersize,journal]{IEEEtran}
\usepackage{amsmath,amsfonts}
\usepackage{algorithmic}
\usepackage{algorithm}
\usepackage{array}
\usepackage[caption=false,font=normalsize,labelfont=sf,textfont=sf]{subfig}
\usepackage{textcomp}
\usepackage{stfloats}
\usepackage{url}
\usepackage{verbatim}
\usepackage{graphicx}
\usepackage{cite}
\usepackage{booktabs}    % \toprule, \midrule, \bottomrule
\usepackage{makecell}    % Multi-line table headers
\usepackage{color} \usepackage[colorlinks=true,citecolor=blue,linkcolor=blue,urlcolor=blue]{hyperref}

\begin{document}

\title{CASAband: Easy-to-Wear Textile Wristband using Shape Memory Alloy Actuators for Spatial and Temporal Haptic Feedback}

\author{Baekgyeom Kim†, Anoush Sepehri†, Jessica Healey, Taeuk Oh, Hyungseok Seo, \\Je-Sung Koh*, Tania K. Morimoto*

\thanks{† These authors contributed equally to this work} 
\thanks{* Corresponding authors: Je-Sung Koh and Tania K. Morimoto}
\thanks{Baekgyeom Kim is with the Korea National University of Transportation, Chungju-si, 27469, South Korea}
\thanks{Anoush Sepehri, Jessica Healey and Tania K. Morimoto are with the University of California, San Diego, La Jolla CA 92093, USA}
\thanks{Taeuk Oh is with Ajou University, Suwon-si, 16499, South Korea}
\thanks{Hyungseok Seo and Je-sung Koh are with Pohang University of Science and Technology (POSTECH), Pohang-si, 37673, South Korea}}

% The paper headers
%\markboth{Journal of \LaTeX\ Class Files,~Vol.~XX, No.~X, August~2021}%
%{Shell \MakeLowercase{\textit{et al.}}: A Sample Article Using IEEEtran.cls for IEEE Journals}

%\IEEEpubid{0000--0000/00\$00.00~\copyright~2021 IEEE}
% Remember, if you use this you must call \IEEEpubidadjcol in the second
% column for its text to clear the IEEEpubid mark.

\maketitle

\begin{abstract}
Haptic interfaces for the wrist and forearm offer an attractive alternative to hand-worn devices as they are simple to wear, leave the hands free for interaction with the real world, and interfere minimally with natural arm motions. To be useful in real-world settings, however, such devices must balance functionality, wearability and comfort, all while being fully untethered with minimal mass and volume. In this work, we present CASAband, a haptic wristband that integrates compliant amplified shape memory alloy actuators (CASA) into a multi-layered textile wristband to deliver spatial and temporal haptic feedback. CASAband operates completely untethered, generates no noise, and has a total mass of 63 g. The device incorporates four actuators that can generate up to 1.7 N of blocked force and 3.2 mm of free displacement with an operating bandwidth ranging from 1.34-6.59 Hz depending on the applied voltage. We conducted a perceptual study and determined that users could identify the location of a single haptic cue around their wrist and discriminate among several patterned cues with over 90\% accuracy on average, highlighting that CASAband can be a suitable wearable interface to deliver information for real-world guidance and navigation tasks. To highlight the potential use cases for CASAband, we conducted two demonstrations: a pick and place task where the user relied only on haptic communication from a moderator, and an outdoor pedestrian navigation task where the user relied only on directional cues on the wrist. CASAband is one of the first haptic interfaces that balances the tradeoff between form and function and presents new opportunities for haptic feedback in the real world.

\end{abstract}


\section{Introduction}
%storyline
Haptic interfaces, which deliver controlled forces and displacements to the body, have been extensively studied for a number of purposes, such as enhancing immersion in virtual environments (VR) \cite{kuchenbecker2006improving,yu2019skin}, supporting or substituting visual and auditory sensation \cite{visell2009tactile}, facilitating human-robot interaction and teleoperation \cite{okamura2009haptic, pacchierotti2023cutaneous}, and providing general guidance for tasks such as surgical robotic training \cite{oquendo2024haptic} or outdoor navigation~\cite{kappers2024hands}. Because mechanoreceptors that respond to tactile sensation are distributed throughout the entire body, haptic devices can be attached to the hands, arms, legs, waist, and torso~\cite{fleck2025wearable}. The majority of existing haptic devices, however, have focused on the hands, where the tactile innervation density is among the highest in the body, and tactile sensitivity is particularly acute \cite{corniani2020tactile}. Typical examples include systems that are mounted directly to the fingertips \cite{giraud2021haptigami, ha2025full, ji2021untethered}, or onto the back of the hand and around the fingers \cite{palagi2023mechanical,xiong2022so, choi2017grabity} to generate cutaneous and kinesthetic haptic feedback. While haptic interfaces on the hands and fingers can deliver immersive haptic feedback, they often prevent the user from interacting with the environment around them, resulting in limited use cases outside of virtual reality. % Consequently, haptic devices worn on the hands and fingers have limited use cases in scenarios outside of VR applications.

\begin{figure}[t]
\includegraphics[width=\linewidth]{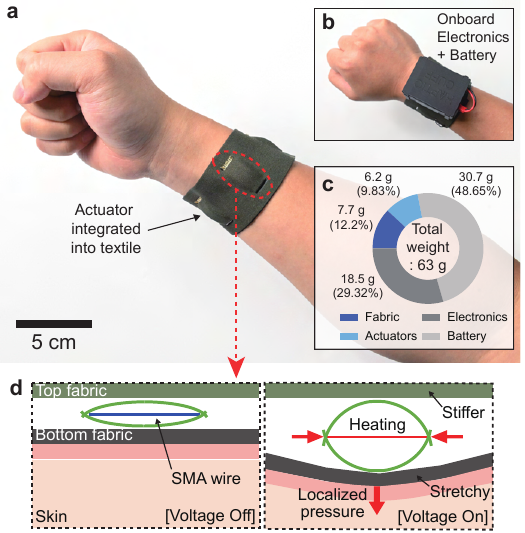}
\vspace{-15pt}
\caption{An overview of the CASAband. (\textbf{a}) The CASAband without any of the onboard electronics. (\textbf{b}) The CASAband with all onboard electronics for untethered operation. (\textbf{c}) Components and weight ratios of the CASAband. (\textbf{d}) Operating principle of the SMA actuators used in the CASAband.}
\label{fig1}
\vspace{-15pt}
\end{figure}

The wrist and forearm, which still have a high density of mechanoreceptors, are promising alternative regions for wearable haptic interfaces. In addition to minimizing the obstruction of the hands, wristbands and bracelets with integrated electronics are widespread and are widely accepted in society \cite{kim2015acceptance}. As a result, researchers have begun to explore techniques to relocate haptic feedback from the fingers to the wrist~\cite{palmer2022haptic, moriyama2022wearable} for applications such as minimally invasive surgery \cite{vuong2025effects} and hands-free navigation \cite{kappers2024hands}. Because the wrist and forearm provide a large region for haptic feedback, researchers have also explored methods to convey information on social touch, such as simulating a stroking sensation by varying the spatial and temporal properties of the haptic cues~\cite{sousa2024evaluating, culbertson2018social, du2024haptiknit}.

\subsection{Existing Actuation Methods for Haptic Interfaces}
Electromagnetic (EM) actuators, such as DC motors and servomotors, have been widely used in haptic devices because they can achieve high forces over a wide operating frequency and can be accurately controlled for either force or displacement. For many motor-based systems, however, the use of mechanical linkages, gears, and guide mechanisms to transmit the desired force and motion to the skin tends to increase the overall size and weight of the device, resulting in systems that can reach several hundred grams (see Table \ref{tab:comparison}) \cite{pezent2022design, meli2018hbracelet, yoshida2024design, moriyama2022wearable}. Not only do these wearable devices have rigid components that can become uncomfortable during longer periods of wear, such weight may be burdensome in everyday use \cite{fleck2025wearable}.

Linear resonant actuators,  voice coil actuators,  and eccentric rotating mass actuators are also common EM actuation methods and have been successfully integrated into many wearable interfaces for cutaneous haptic feedback \cite{pezent2022design, culbertson2018social}. Eccentric rotating mass actuators are inexpensive and capable of providing vibration feedback, however, they are driven by a single voltage input that couples frequency and amplitude \cite{fleck2025wearable}. Linear resonant actuators and voice coil actuators offer independent control over the frequency and amplitude of the vibration but are most effective for generating high frequency tactile feedback $(>100~hz)$ which only excites the Pacinian corpuscle receptors in the body \cite{culbertson2018haptics}. Existing literature has shown that vibration feedback is traditionally limited to conveying binary information \cite{culbertson2018haptics} and has poor spatial acuity because of the low density of Pacinian corpuscle receptors and propagation of vibration through the soft tissue \cite{sofia2013mechanical}. Furthermore, extended vibration can be interpreted as unpleasant by the user \cite{sanchez2024cutaneous} and can lead to a tingling or numbing sensation even after the stimuli is removed \cite{ribot1996alteration}. 

Electrostatic soft actuators, such as dielectric elastomer actuators (DEA), and electrohydraulic actuator, such as HASEL actuators, offer high energy densities and broad operating bandwidths in a thin, compliant body, and have been actively investigated for haptic interfaces \cite{zhao2020wearable, lee2022wearable, sanchez2024cutaneous}. While highly responsive, the limited force (\textless 0.6 N) and stroke (0.3~mm)  of DEAs may act as a constraint when high-intensity haptic feedback is needed \cite{zhao2020wearable, lee2022wearable}. Electrohydraulic actuators utilizing fluid motion inside soft pouches can generate higher forces and displacements by using a multilayer pouch structure \cite{sanchez2024cutaneous}. Nonetheless, the complex fabrication process and the requirement for a high-voltage driving circuit remain a common challenge, particularly when extending such systems toward long-term, fully wearable haptic systems with multiple inputs for spatial feedback.

Pneumatic actuators are widely used in soft haptic devices because of their favorable force and displacement properties and conformal structure capable of adapting to the skin \cite{young2019bellowband, jumet2022textile, zhu2020pneusleeve, he2015pneuhaptic, du2024haptiknit, zhang2024haptic}. These studies suggest that pneumatic actuators can achieve both high force output and be comfortable to wear when integrated into a textile-based wearable. However, pneumatic actuators require a source of pressurized fluid, typically from a pump or a reservoir, as well as valves which increase the weight, complexity, and noise in the system, again limiting their untethered use. Recent work has investigated combining multiple pneumatic cells with fluidic circuits to generate various haptic cues and spatiotemporal patterns using a single pressure input \cite{jumet2023fluidically}. While this successfully reduced the number of necessary valves to operate the device, the haptic cues were limited to predefined patterns and users were still required to wear a compressed air canister and electrical components on their waist. 

Thermal actuators, such as shape memory alloy (SMA) and liquid crystal elastomers (LCE), have a high force-to-weight ratio, and do not inherently require large backend equipment or specialized high voltage modules during operation. As a result, thermal actuators are an attractive option for lightweight and portable wearable haptic interfaces\cite{sousa2024evaluating, hamdan2019springlets, gupta2017hapticclench, sepehri2025bundled, oh2023easy, forman2023fiberobo}. While wrist- and forearm-worn devices employing thermal actuators can generate forces on the order of 1-10 N depending on the size of the actuator, their operating frequencies are typically well below 0.33 Hz (see Table \ref{tab:comparison}). While some work introduced bundling techniques with active cooling to minimize the tradeoff between the force generated and the response speed to achieve bandwidths above 1 Hz, this increased the complexity of the fabrication procedure and necessary backend components during operation \cite{sepehri2025bundled}. Moreover, poor controllability due to the nonlinearity and hysteresis of the material, as well as low energy efficiency have hindered the practical usage of such actuators in wearable devices to date.

\begin{table*}[!t]
    \centering
    \renewcommand{\arraystretch}{1.5}
    \setlength{\tabcolsep}{5pt}
    \caption{Comparison of wristband/armband haptic devices (squeeze and localized pressure)}
    \label{tab:comparison}
    \begin{tabular*}{\textwidth}{@{\extracolsep{\fill}}lccccccc@{}}
        \toprule
        \makecell{} & \makecell{Actuator\\type} & \makecell{Max Force\\(N)} & \makecell{Max Disp.\\(mm)} & \makecell{Max Freq.\\(Hz)} & \makecell{Controllable\\Units} & \makecell{Act. Weight\\(g)} & \makecell{Dev. Weight\\(g)} \\
        \midrule

        \multicolumn{8}{l}{\textbf{Tethered Power Source}} \\
        Culbertson et al.\cite{culbertson2018social} & Voice coil & 2 & 4 & 1 & 6 & 29.0 & — \\
        Pezent et al.\cite{pezent2022design} & \makecell{DC motor\\\& LRA} & 15 & — & 9.1 & 1 & — & 120.0 \\
        Zhao et al.\cite{zhao2020wearable} & DEA & 0.6 & 0.3 & 200 & 4 & 0.6 & — \\
        Young et al.\cite{young2019bellowband} & Pneumatic & 12 & 10 & 7 & 8 & — & 11.0 \\
        Jumet et al.\cite{jumet2022textile} & Pneumatic & 67 & — & — & 1 & — & — \\
        Zhu et al.\cite{zhu2020pneusleeve} & Pneumatic & 1.65 & — & 4 & 6 & — & 26.0 \\
        Gupta et al.\cite{gupta2017hapticclench} & SMA & 15 & — & 0.06 & 1 & — & — \\
        Oh et al.\cite{oh2023easy} & SMA & 1.25 & — & 0.07 & 4 & — & — \\
        Sepehri et al.\cite{sepehri2025bundled} & LCE & 1.2 & 4.1 & 0.91 & 1 & — & — \\
        %Forman et al.\cite{forman2023fiberobo} & LCE & 0.045--0.13 & — & 0.028--0.6 & 1 & — & — \\

        \midrule

        \multicolumn{8}{l}{\textbf{Onboard Electronics \& Power Source}} \\
        Meli et al.\cite{meli2018hbracelet} & \makecell{Servo \& linear\\actuator} & 32 & 13 & — & 5 & — & 306.0 \\
        Lee et al.\cite{lee2022wearable} & DEA & 0.2 & 0.4 & 240 & 15 & 0.6 & 220.0 \\
        He et al.\cite{he2015pneuhaptic} & Pneumatic & — & — & 0.2 & 5 & — & — \\
        du Pasquier et al.\cite{du2024haptiknit} & Pneumatic & 40 & — & 14.5 & 8 & — & 440.0 \\
        Hamdan et al.\cite{hamdan2019springlets} & SMA & 2.4 & — & 0.33 & 1 & — & — \\
        \textbf{Our work} & \textbf{SMA} & \textbf{1.7} & \textbf{3.2} & \textbf{1.34--6.59$^{*}$} & \textbf{4} & \textbf{0.2} & \textbf{63.1} \\

        \bottomrule

        \multicolumn{8}{p{\textwidth}}{\footnotesize
        $^{*}$This value corresponds to the $-3$ dB bandwidth estimated from the stroke--frequency profiles of the actuator in the textile wristband (see Fig.~\ref{fig5}b). The 100 ms actuation condition was used for the user perception study and demonstrations in this work. Additional details for estimating the cutoff frequency are provided in the supplementary material.
        }
    \end{tabular*}
    \vspace{-15pt}

\end{table*}

\subsection{Design Considerations for Wearable Haptic Interfaces}
Regardless of the type of actuator used, the interface to anchor the actuator and transmit the force to the body considerably influences the overall success and adoption of the final device. Traditionally, haptic devices have used rigid components such as plastics and metals to anchor onto the body \cite{yoshida2024design, moriyama2022wearable}. Unfortunately, this approach can increase the bulkiness of the device and reduce comfort and wearability, ultimately limiting daily use. Textiles offer a promising medium for wearable devices given they are lightweight, flexible, and have similar stiffness and mechanical properties to the skin \cite{banerjee2018soft}. Moreover, a variety of wearable structures can be fabricated using techniques such as cutting, layering, laminating, and customized embroidery or knitting \cite{sepehri2025retrofitting, jumet2022textile, du2024haptiknit}, to easily create garment-like devices that are familiar to the user.

Several considerations should be addressed in parallel when developing a haptic interface for daily use in the real world. First, the device should achieve sufficient mechanical performance in terms of force, displacement, and speed in order to be capable of generating a variety of cues and patterns that are easily discernible to the user. Second, wearability is essential for long-term daily use, necessitating a device that is small, lightweight, silent, and comfortable to avoid impairing the user’s natural movements \cite{yin2021wearable}. Lastly, to support unrestricted activity in real-world environments, the system should be untethered with onboard electronics and power. Although several existing haptic devices satisfy some of these requirements, the weight of the total system when onboard electronics and batteries are included often reaches hundreds of grams (see Table \ref{tab:comparison}) which significantly hinders use in real-world settings. Developing a wearable haptic interface for the real world requires a design approach that jointly considers performance, form factor, and integration of onboard electronics and power which is an open challenge to date.

\subsection{Contributions}
In this work, we present CASAband: a lightweight, untethered haptic interface that integrates compliant amplified SMA actuators (CASA) into a textile wristband to provide spatial and temporal haptic feedback for real-world guidance and navigation applications (Fig.~\ref{fig1}). Compared to existing thermally-actuated haptic interfaces, CASAband is capable of generating a variety of spatial and temporal haptic cues, rather than just relying on binary haptic feedback, is responsive at frequencies over 1~Hz, and is capable of operating for over 4 hours on a single battery charge. Furthermore, the entire device occupies a small form factor, requires minimal operational equipment, and is designed in the form of a textile wristband to maximize accessibility and comfort.

To develop CASAband, we first designed a compact, lightweight, and high-powered SMA actuator that achieved a blocked force of 1.7 N and free displacement of 3.2~mm at frequencies well over 1 Hz. We then integrated the actuator into a multi-layered textile wristband that weighed 63~g (including the battery and all electronic components) to provide spatial and temporal haptic feedback to the wrist while being completely noise-free and untethered. We conducted a user study and determined that participants could localize different haptic cues around the wrist and discriminate between seven haptic patterns that varied in speed, sequence, and magnitude, achieving accuracies over 90$\%$ on average for all experiments. Finally, we demonstrated the technical readiness and capabilities of CASAband in two practical use-case scenarios: (1) a pick and place demonstration where a user, with no visual or auditory information, had to rely solely on haptic feedback for instructions to complete a task, and (2) a pedestrian navigation demonstration where a user was guided through a complex route with multiple waypoints by only relying on commands from the CASAband. To the best of our knowledge, CASAband is the first fully untethered thermally-actuated wristband that is capable of generating a variety of spatial and temporal haptic cues suitable for real-world guidance and navigation tasks. 

\section{Materials and Methods}

\begin{figure*}
\includegraphics[width=\linewidth]{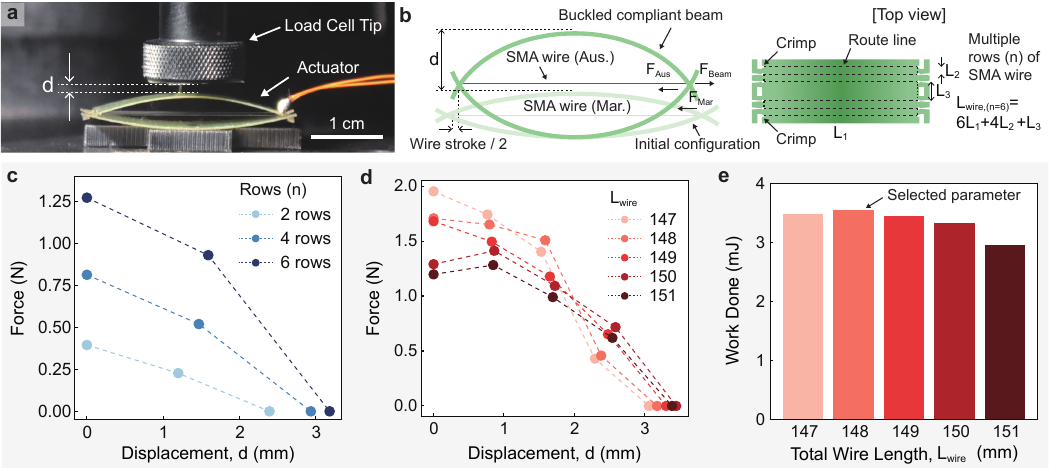}
\vspace{-15pt}
\caption{Characterization of the SMA actuator used in the CASAband. (\textbf{a}) Experimental setup for measuring the blocked force at different displacements, d. (\textbf{b}) Actuation mechanism and design parameters of the SMA actuator. (\textbf{c}) Measured blocked force when varying the number of rows of SMA wire. (\textbf{d}) Measured blocked force when varying the total SMA wire length for 6 rows. (\textbf{e}) The calculated work done by the actuators tested in (d).}
\label{fig2}
\vspace{-15pt}
\end{figure*}

\subsection{Design Requirements}

We surveyed existing literature to establish functional design requirements for CASAband to ensure we could generate a diverse set of cues and patterns that would be easily interpretable by the wearer. Specifically, we focused on establishing design requirements for the displacement, force, and bandwidth of the actuators, as well as the spacing around the wrist.  

As a haptic device applies forces to the skin, the soft tissue deforms, stimulating the mechanoreceptors in the body. As a result, researchers have investigated the minimum detectable thresholds for touch in terms of both displacement and force. For displacement, existing work suggested that a normal indentation displacement of 1.5 mm on the forearm was the smallest signal that could be confidently detected by humans  ~\cite{biggs2002tangential}. For force, existing haptic interfaces that used dielectric elastomer actuators~\cite{zhao2020wearable} or electromagnetic actuators~\cite{erwin2014design} generated up to 0.6~N and resulted in over 80\% accuracy in localization and pattern identification. Furthermore, Kodali et. al. \cite{kodali2023wearable} determined that the minimum perceptible threshold for contact pressure on the forearm was 0.41~N. Based on these results, we aimed to design our actuator to achieve over 1~N of force at 1.5~mm of displacement to ensure we could generate clear haptic cues to the user. 

To develop a haptic device for guidance and navigation assistance, the haptic feedback must be fast enough to respond to real-world events (e.g., heading corrections during walking). Unlike applications such as surface or texture rendering, which require bandwidths in the tens to hundreds of hertz, guidance and navigation tasks can be effectively supported with update rates of 1 Hz for environment and task level information and corresponding haptic feedback rendered every few seconds \cite{van2004waypoint} or slowly over the span of several seconds \cite{ploch2017comparing}. At these lower frequencies, haptic feedback is primarily delivered through localized pressure and sustained skin deformation, which are detected by the Merkel disks and Ruffini corpuscles mechanoreceptors. These mechanoreceptors are particularly sensitive to low-frequency stimuli on the order of ~1 Hz \cite{fleck2025wearable, saal2014touch}.  Based on these findings, we aimed to achieve a bandwidth of at least 1 Hz, while prioritizing the force and displacement of our actuator to maximize the variety of haptic cues we could generate.

Existing literature suggested that the two point discrimination on the arm and forearm ranged from 30.7~mm to 45.4~mm~\cite{nolan1982two}. As a result, we placed each actuator approximately 35~mm and 40~mm apart in a small and large version of our textile wristband for different wrist circumferences to achieve varying spatial haptic feedback around the wrist. Further reducing the spacing between the actuators would increase the complexity of our device without considerably improving the haptic experience.

\subsection{Compliant Amplified SMA Actuator (CASA) Design}

In our previous work, we developed a compliant amplified SMA actuator (CASA) with a high power-to-weight ratio (1.7~kW/kg) and thin form factor to apply localized pressure feedback on the skin ~\cite{kim2022actuating}. The CASA consisted of two compliant beams (200 $\mu m$ GFRP, Sungsim Tech) for the strain amplification mechanism and the SMA wire (Dynalloy Co., USA) (see Fig.~\ref{fig2}a and Fig.~\ref{fig2}b). The initial shape of the CASA was an elliptic configuration, and the compliant beam amplified the horizontal contraction of the SMA wire (typically 3-5\% \cite{zhang2019robotic}) into vertical displacement through structural deformation. These actuators are particularly well-suited for wearable devices that are subject to strict volume and weight constraints. In our previous work, users were able to distinguish different magnitudes of force in both continuous and impulsive stimuli on the hand, which confirmed the feasibility of employing such actuator in a wearable haptic device \cite{kim2022actuating}. 

We performed characterization experiments to tune the design parameters of the CASA in order to satisfy our performance requirements. The CASA had three design parameters: the diameter of the SMA wire, the number of rows of the embedded SMA wire, and the length ratio between the compliant beam and SMA wire. These design parameters influenced the operating frequency, force, and displacement of the final actuator. To achieve a bandwidth over 1 Hz,  we selected a diameter of 0.05~mm for the SMA wire. The cooling time for this diameter provided by the manufacturer was~0.3 s, which is faster than many other thermally driven actuators and comparable to some pneumatic devices (see Table~\ref{tab:comparison}). 

To measure the blocked force of the CASA at different displacements, we used a tensile testing machine (Instron 3343) with the load cell tip placed at fixed distances from the CASA (see Fig.~\ref{fig2}a).  To compare actuator performance under equivalent conditions, each actuator was driven with the same electric current (85~mA) and actuation time (1~s) provided by the manufacturer's guidelines (Dynalloy Co., USA). This procedure was necessary because the length of the SMA wire varied with the actuator design parameters, which would result in different electric currents even under the same applied voltage. For all experiments performed after selecting the design parameters of the actuator, we described the electrical input in terms of the applied voltage and actuation time.

For normal stimulation on the forearm, the target actuation force was 1 N at 1.5~mm displacement. Although thinner SMA wires have faster cooling rates, and therefore higher bandwidths, the force generated by the wire decreases significantly because it is directly proportional to the cross sectional area. To design an actuator capable of generating the target force, we used multiple rows of the SMA wire and conducted experiments to measure the resulting blocked force (see Fig.~\ref{fig2}c). Each actuator routed with different rows ($n=2,4,$ and $6$) was designed to have a consistent initial height of about 5.5 mm. This ensured that the proportional increase in blocked force was solely due to the variation in row count and not the initial height. Based on our results, we selected six rows to satisfy the target specification.

The actuator’s force-displacement relationship was also related to the length ratio between the compliant beam and SMA wire \cite{kim2022actuating}. Therefore, to design an actuator that can achieve optimal actuation performance under the selected parameters (diameter and rows), we obtained the force-displacement relationship according to the total wire length (see Fig.~\ref{fig2}d). The mechanical work done by the actuator, as calculated by the area under the force-displacement curves, exhibited a local maximum for 148~mm of total wire length (see Fig.~\ref{fig2}e).

The initial height of the actuator was an important factor that could influence wearability, as it determined the overall thickness of the final wristband. The initial height of the actuator ranged from 5.87 mm to 4.13 mm for $L_{wire}$ of 147 mm to 151 m (see Fig.~S1). To deliver clear tactile feedback on the forearm, we prioritized maximizing the force and displacement properties of the actuator. While the height increased marginally when optimizing for the force and displacement ($\sim$1.7 mm), this difference was not considered critical for the overall form factor of the device. Although the selected parameters do not allow the minimum possible thickness, the resulting height was still considerably smaller than other soft actuators used for localized pressure feedback ($\sim$1 cm) \cite{sanchez2024cutaneous}. 

\begin{figure}
  \includegraphics[width=\linewidth]{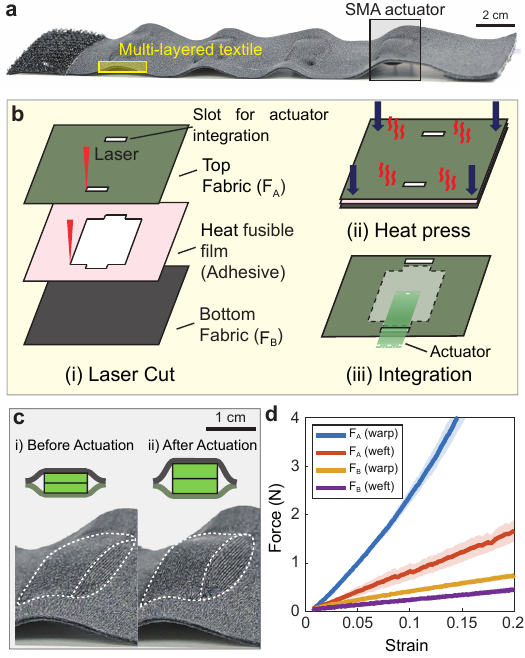}
  \vspace{-15pt}
  \caption{Design and fabrication of the textile wristband for the CASAband. (\textbf{a}) The CASAband with four integrated actuators before donning. (\textbf{b}) Fabrication process of the multi-layered textile. The multi-layered textiles are (i) laser cut to shape, (ii) bonded together using a heat-fusible film, then (iii) integrated with the actuators. \textbf{(c)} Before (i) and after (ii) activating the SMA actuator embedded in the wristband (overlay of the actuator shown for clarity). (\textbf{d}) Force-strain relationship of each fabric by pulling in the warp and weft directions. The top (outer) fabric is considerably stiffer than the bottom (inner) fabric to ground the actuator against the skin. The markers and shaded region in (d) represent the mean and standard deviation respectively (n=3)}
  \label{fig3}
  \vspace{-15pt}
\end{figure}

\subsection{Textile Wristband Integration}

With our selected parameters for the CASA, we integrated four actuators into a comfortable, easy-to-wear wristband made from everyday textiles seen in wearable apparel. The haptic wristband consisted of two textile layers bonded together using a flexible heat fusible film (see Fig.~\ref{fig3}a and Fig.~\ref{fig3}b). A heat fusible fabrication approach eliminated the need for complex cut-and-sew procedures and has already been successfully implemented to develop haptic interfaces~\cite{jumet2022textile} and wearable sensing garments~\cite{sepehri2025retrofitting}. We first laser cut the fabric and heat fusible films to the appropriate shapes using a CO2 laser cutter (see Fig.~\ref{fig3}b(i)). We then bonded the heat fusible film (HeatnBond Soft Stretch, Therm O Web) to the outer fabric using a heat press (4s @ 120~°C). We then removed the protective backing from the heat fusible film and bonded the two fabrics together with the heat press (12s @ 120~°C, see Fig.~\ref{fig3}b(ii)). We aligned the weft of the fabric with the length of the wristband to provide some compliance when wrapping around different user’s wrists. We fabricated two textile wristbands, a small wristband where each actuator was spaced 35~mm apart and a large wristband where each actuator was spaced 40~mm apart to accommodate different users. We then integrated the CASA into the wristband so that each actuator was equidistant around the wrist (see Fig.~\ref{fig3}b(iii) and Fig.~\ref{fig3}c). To maximize the force transmitted from the actuator to the skin, we selected an outer fabric that was considerably stiffer than the inner fabric (see Fig.~\ref{fig3}d). This approach was similar to previous work that developed distributed stiffness garments for pneumatically actuated haptic interfaces~\cite{du2024haptiknit}. By using off-the-shelf textiles and assembling them using a 2D heat lamination strategy, however, we can achieve similar results without the need for custom knitting equipment and hardware.

\begin{figure*}
  \includegraphics[width=\linewidth]{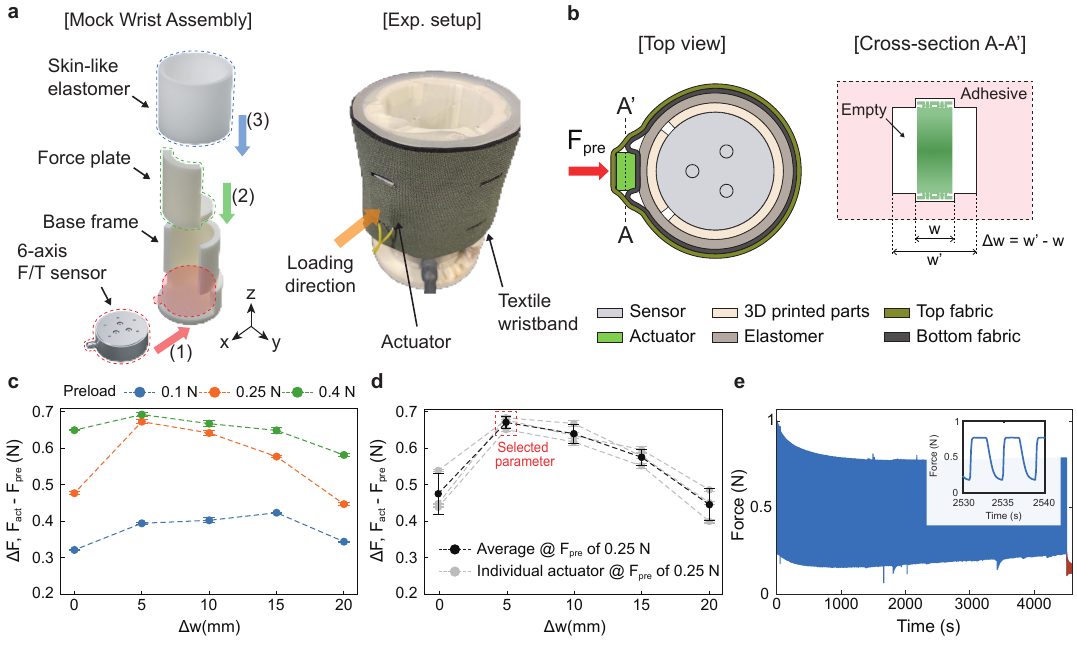}
  \vspace{-15pt}
  \caption{Characterization results for the CASAband. (\textbf{a}) Experimental setup composed of the instrumented mock wrist and the CASAband. (\textbf{b}) Schematic of the cross-section showing the direction of the preload. Parameter $\Delta w$ was defined as the gap tolerance between the heat fusible film and the actuator. (\textbf{c}) Relationship between the force output and $\Delta w$ under different preload conditions. (\textbf{d}) Experimental results for three SMA actuators using the optimized preload and $\Delta w$. (\textbf{e}) Life cycle assessment of the actuators used in the CASAband. The actuator can survive over 1000 cycles of continuous operation. The data point and error bars in (c) and (d) represent the mean and standard deviation respectively (n=3).}
  \label{fig4}
  \vspace{-15pt}
\end{figure*}
 
\subsection{Onboard Electronics for Untethered Operation}
CASAband consisted of a textile wristband with four integrated CASAs that were independently operated via joule heating. In order to achieve the untethered operation seen in the demonstrations described in Section~V, we integrated all electronic components, such as wireless communication capabilities, motion sensing using an inertial measurement unit (IMU), and an onboard power, all into a module that is directly attached to the wristband (see Fig.~\ref{fig1}b). 

The electronics unit included a development board (Arduino Nano 33 BLE) for wireless communication, an IMU sensor (LSM9DS1) for user gesture recognition, two 12 V step-up regulators (U3V16F12, Pololu) for actuator operation, a  5V step-down for logic power (D45V5F5, Pololu), two dual motor drivers (TB6612FNG, Pololu), and a high discharge rate lithium polymer (LiPo) battery (7.4V, 350mAh, 20 C-rate). The electronic unit weighed 49.2~g, including the enclosure, while the entire device weighed 63~g (Fig.~\ref{fig1}c).

In addition to its compact and lightweight design, CASAband can operate for long periods on a single charge in real-world outdoor environments. We evaluated the operating time of the device, which was battery-powered and received commands from a smartphone application via Bluetooth. The device sequentially generated the haptic cues and patterns used in the user perception study (see Section~IV) and demonstrations (Section~V) at 10-second intervals (see supplementary video). Under these conditions, the device operated continuously for over four hours on a single charge. Considering that the average interval between cues for the demonstrations in Section~V were 7 and 11 seconds, respectively, this battery life confirms that CASAband can be operated for several hours on a single charge.

\section{CASAband Characterization Results}

\subsection{Force Characterization}
We measured the force generated from the CASAband on an instrumented test rig to simulate the forearm. The test rig consisted of a 6-DOF force/torque sensor that was integrated into a cylindrical structure wrapped with a layer of silicone to simulate the compliance of the forearm and wrist of a user (see Fig.~\ref{fig4}a). We 3D printed the mock forearm test rig using Polylactic Acid (Bambu X1, Bambu Labs). The tool side was mounted directly to the 6-DOF force/torque sensor (Mini40, ATI Industrial Automation). We incorporated a 4 mm-thick layer of silicone (Ecoflex 00-30, Smooth-On) to replicate the compliance of soft tissue based on existing literature \cite{sparks2015use}. To control the preload on the actuator, we manually wrapped the CASAband around the test rig while monitoring the force readings from the sensor in order to tighten and loosen accordingly.

When fabricating the wristband, we observed that the width of the heat fusible film changed the effective stiffness of the top and bottom textiles, therefore changing the amount of force transmitted to the user. We also observed that varying the tension of the wristband when donned impacted the amount of force generated by the actuators, likely due to the preload on the actuator (see Fig.~\ref{fig4}b). As a result, we evaluated the force generated by the actuator in the textile wristband when varying both the preload and the width of the heat fusible film using our instrumented test rig.

We activated the actuators at 8~V for 2 seconds and recorded the maximum force generated. We conducted each experiment under different preload conditions ranging from 0.1~N to 0.4~N (see Fig.~\ref{fig4}c and Fig.~\ref{fig4}d). While larger preloads (i.e., tightening the wristband) could generate larger forces, too high of a preload caused the compliant mechanism in the CASA to buckle and fail prematurely. For the 0.25~N and 0.4~N preload conditions, we observed that the design with the 5~mm wide adhesive resulted in the highest force. We hypothesized that an adhesive that was too narrow resulted in too high of a stiffness, therefore causing the actuator to buckle. An adhesive that was too wide, however, would not provide enough stiffness to reliably transmit the force into the soft tissue. Based on our results, we selected the 5~mm wide adhesive and aimed for at least 0.25~N of preload for the remaining experiments. We achieved a similar preload by qualitatively monitoring the tension when wrapping the CASAband around the user's wrist. Because the force generated by our device was similar for preloads ranging from 0.25~N to 0.4~N, we believe any variations in tension when donning the device had minimal impact on the quality of the haptic feedback.

\subsection{Durability Characterization}

We used the mock forearm test rig to evaluate the lifespan of a single actuator in the CASAband. We operated the actuator at 8 V and 0.25 Hz (2 seconds on, 2 seconds off) and collected data from the F/T sensor continuously at 10 Hz.

The actuator in the CASAband was capable of achieving 1127 cycles at maximum force before failing (see Fig.~\ref{fig4}e). While we initially tightened the textile wristband to achieve a preload of approximately 0.25 N, the preload decreased and settled after approximately 300 cycles, most likely due to small movements of the fabric on the test rig during the initial actuation cycles. At the point of failure, the SMA wire became damaged, therefore leading to non-uniform heating that eventually led to the wire snapping. Nevertheless, CASAband was capable of achieving over 1000 cycles in sustained operating conditions, therefore making it suitable for real-world applications.

\begin{figure}
  \includegraphics[width=\linewidth]{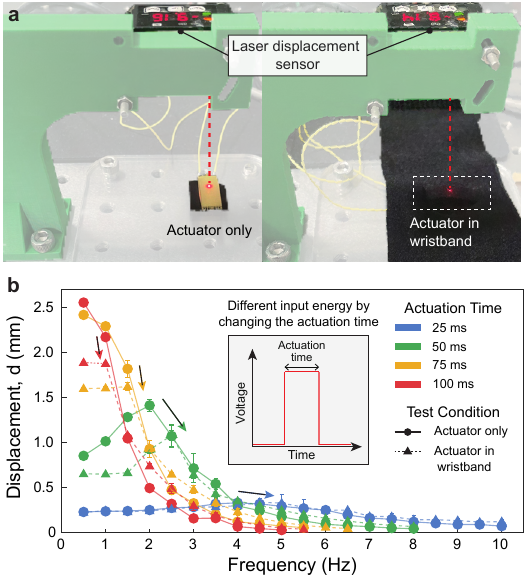}
  \vspace{-15pt}
  \caption{Dynamic characterization results for the CASAband. (\textbf{a}) Experimental setup for measuring the actuator’s displacement with and without the textile wristband. (\textbf{b}) Frequency characteristic results by varying the time the actuator was on. The data point and error bars represent the mean and standard deviation respectively (n=3).}
  \label{fig5}
  \vspace{-15pt}
\end{figure}

\subsection{Operating Bandwidth Characterization}

To provide users with low-latency haptic cues, the actuators in a haptic devices should have a sufficiently fast response times for the given application. We measured the displacement of the actuator both inside and outside of the textile wristband under different operating conditions. The experimental setup consisted of a laser displacement sensor (Panasonic HG-C1030-P) fixed to a 3D printed jig to measure the actuator's vertical displacement. The laser displacement data were acquired using a DAQ (Data Acquisition) system (DEWESoft, Ltd., SIRIUS) (see Fig.~\ref{fig5}a).

When a 12~V signal was applied for 100~ms, the SMA wire in the actuator reached approximately 90 °C, the theoretical completion point of its Austenite phase transition (see Supporting Information for the estimation). We varied the actuation time from 25 ms to 100 ms, while maintaining a constant voltage input (12~V), to control the input energy and evaluate the bandwidth of the actuator. For the evaluated actuation times, the bandwidth ranged from 1.34 Hz for the 100 ms actuation time up to 6.59 Hz for the 25 ms actuation time respectively (see Fig.~\ref{fig5}b). 

When outside of the textile wristband, the displacement of the CASA decreased with increasing frequency under the 75-ms and 100-ms actuation conditions. This behavior was attributed to an insufficient cooling time between cycles, causing the displacement of the actuator to saturate. Under shorter actuation times (25-ms and 50-ms), the displacement was smaller at low frequencies since the actuator did not achieve full Austenite phase transition and fully cooled before the next heating cycle. We observed, however, peak displacement at specific frequencies: 2 Hz for the 50 ms actuation time and 5 Hz for the 25 ms actuation time, suggesting that there was a balance point between cyclic low-energy input and the cooling time at which the displacement (and thus bandwidth) was maximized. We believed this was due to the buildup of heat which caused the actuator to operate within a larger region of the phase transition temperatures after multiple actuation cycles. Note that these resonance-like peaks are distinct from mechanical resonances governed by system mass and stiffness; instead, they arise from the interplay between cyclic heating and the actuator’s thermal recovery (cooling) dynamics.

When embedded in the textile wristband, the actuator exhibited a similar overall trend. Notably, displacements were maintained in the low-frequency range (below 2 Hz), presumably because the stretchable fabric of the wristband acted as an additional force to help restore the shape of the CASA when at room temperature. Prior work has shown that a bias force or pre-stress can affect the strain rate of elongation during the cooling process, which can consequently influence the actuation frequency \cite{tadesse2010tailoring,liu2020actuation}). Although the displacement of the actuator embedded in the wristband decreased in the low-frequency range, it remained comparable or even higher than that of the free actuator at specific frequencies. 

In summary, the CASA integrated into the textile wristband showed a rapid response ($<$100 ms to reach maximum stroke) with a bandwidth over 1 Hz. Compared to other thermally driven actuators, CASAband achieved a higher operating frequency while maintaining a comparable actuation force (see Table \ref{tab:comparison}).

\subsection{Operating Temperature Characterization}

\begin{figure}
  \includegraphics[width=\linewidth]{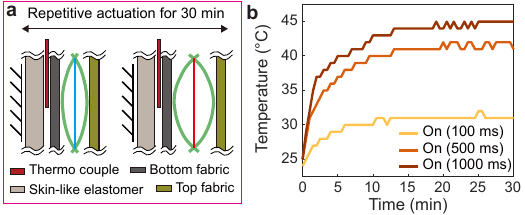}
  \vspace{-15pt}
  \caption{Temperature characterization results. \textbf{(a)} Experimental setup. The thermocouple was placed in between the elastomer on the instrumented test rig and the actuator in the CASAband. \textbf{(b)} Contact temperature on the skin-like elastomer after continuous operation for 30 minutes with actuation times ranging from 100 ms to 1000 ms. For all actuation conditions the cooling time was fixed at 1000 ms between each cycle.}
  \label{fig6}
  \vspace{-15pt}
\end{figure}

We performed a thermal characterization test to ensure that the heat generated during continuous operation did not result in uncomfortable temperatures on the skin. Although the SMA wire must operate at over 90 °C to achieve its full Austenite phase transition, the SMA wire does not come in direct contact with the skin. Furthermore, the compliant beam and textile wristband acted as an insulating barrier to protect the user. We measured temperature by placing a thermocouple (TL0201, PerfectPrime) between the CASAband and the silicone layer of the instrumented forearm test rig (See Fig.~\ref{fig6}a). The actuation conditions were set to 100 ms on at 12~V, followed by 1 second to cool. The temperature was measured for 30 minutes with repeated operation, and converged between 31 and 32 degrees Celsius (See Fig.~\ref{fig6}b). This temperature is safe even when in direct contact with the skin, and is within the thermal comfort zone for the human forearm \cite{yao2007experimental}. We performed an additional temperature measurement experiment for 30 minutes under more extreme conditions (1 second on at 8 V, followed by 1 second off). The measured temperature did not exceed 45 degrees; however, continuous actuation at temperatures above 40 degrees to the skin can be noticeable to the user and potentially uncomfortable. To minimize the accumulation of heat that could be noticed by the user, we limited the duration of each haptic pattern in the user study and demonstrations to several seconds. Furthermore, while the temperature characterization tests were conducted with 1-second intervals between actuation cycles, practical use cases (e.g., conveying task instructions and GPS navigation: see Section 2.7) involved substantially longer intervals between cues, averaging 7 and 11 seconds respectively. This ensured that that the accumulation of heat in real-world settings was even lower than our experimental measurements.  

\begin{figure*}
\includegraphics[width=\linewidth]{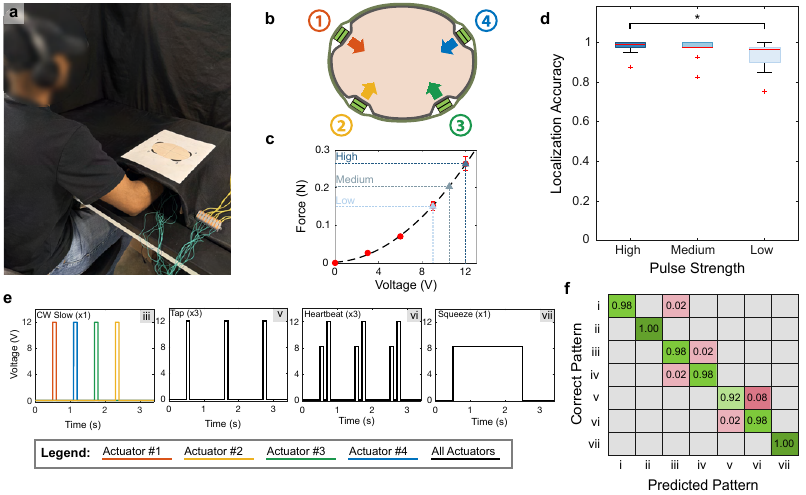}
\vspace{-15pt}
\caption{Results for the user perception study. (\textbf{a}) Experimental Setup. (\textbf{b}) Each CASA was equally spaced around the user’s wrist and labeled 1-4 for the user to identify. (\textbf{c}) Force characterization results on the mock forearm test rig for 100 ms pulse inputs at different voltages. High, medium, and low cues for the user study were selected based on pilot user results. (\textbf{d}) Localization performance results. Participants could detect the location of single haptic cues around their wrist with over 93$\%$ accuracy for all voltages tested. (\textbf{e}) Representative haptic patterns selected for the pattern identification study. (\textbf{f}) Confusion matrix summarizing the pattern identification results. Users could distinguish between all patterns with over 92$\%$ accuracy.}
\label{fig7}
\vspace{-15pt}
\end{figure*}

\section{User Perception Study}
We conducted two haptic perception studies with 10 participants to evaluate the variety of haptic cues and patterns that could be successfully rendered and identified by the user.

\subsection{User Study Procedure}

This human-subject study was approved by the Institutional Review Board at the University of California, San Diego. The study consisted of 10 participants (9 right-handed, 1 left-handed; 7 male, 3 female), and all participants gave informed consent. First, we secured the large wristband on the participant; if the Velcro region overlapped with any of the actuators, the smaller wristband was used instead. The participants wore noise-canceling headphones to minimize audible disturbances, and the wristband was worn on their dominant hand and kept out of view to eliminate any visual aids that could help with the study (see Fig.~\ref{fig7}a).

The study consisted of two phases: (1) the localization identification phase and (2) the pattern identification phase. For the first phase, we generated 12 different cues, corresponding to the 4 locations around the wrist at three different voltage inputs. The voltage levels were selected based on pilot experiments, where we determined that a fast cue (12~V for 100~ms) provided the strongest sensation. We then characterized the force output using our mock forearm test rig for 100~ms pulses with varying voltage inputs to select the low, medium, and high cue (see Fig.~\ref{fig7}c). Finally, we selected the high voltage based on the maximum safe operating conditions for the actuator and the low voltage based on the minimum perceptible threshold we observed during preliminary experiments. We repeated each cue 10 times for a total of 120 cues per participant. While we did vary the location and the magnitude of the cue at the same time, the participant was only tasked with identifying the location.

For the second phase, we generated seven different patterns that we designed specifically for haptic guidance applications. Each pattern is described below:

\textbf{(i) CCW slow:} a slow counterclockwise rotation cue generated by sequentially activating each actuator at 12 V for 100 ms with a 500 ms delay between actuators.

\textbf{(ii) CCW fast:} a fast counterclockwise rotation cue generated by reducing the delay between actuators from 500 ms to 100 ms.

\textbf{(iii) CW slow:} a slow clockwise rotation cue generated by following the same timing sequence as (i), but reversing the direction of actuation. We started each pattern (i)-(iv) using actuator 1 (see Fig.~\ref{fig7}b) to prevent participants from distinguishing between the clockwise and counterclockwise patterns based on the starting location (see Fig.~\ref{fig7}e(iii)).

\textbf{(iv) CW fast:} a fast clockwise rotation cue with the same timing sequence as (ii) but with the actuation order from (iii)).

\textbf{(v) Tap:} a tapping cue generated by activating all actuators at 12 V for 100 ms with a 1000 ms delay (see Fig.~\ref{fig7}e(v)).

\textbf{(vi) Heart beat:} a heart beat cue generated by activating all actuators at 8 V for 100 ms, followed by a 100 ms delay, then immediately activating all actuators at 12 V for 100 ms, and lastly followed by a 700 ms delay (see Fig.~\ref{fig4}e(vi)).

\textbf{(vii) Squeeze:} a squeeze cue generated by activating all actuators at 8 V for 2000 ms followed by a 2000 ms delay (see Fig.~\ref{fig4}e(vii)). 

A single haptic cue consisted of three cycles of the patterns described above. We repeated each cue 5 times for a total of 35 rendered cues for each participant for the study.

Before each phase, participants underwent a practice session to become familiarized with the potential cues that could be generated during the study (up to 7 minutes). During this training, participants were given a visual handout that showed either four numbered regions around the wrist (see Fig.~\ref{fig7}b) or the seven haptic patterns they would feel. Participants were asked to determine the location of the rendered impulsive cue, or to determine the rendered haptic pattern. The order of the cues was randomized, and there were two instances of each cue per practice session. The moderator provided feedback on the correctness of their response. 

Participants then began the formal study (up to 15 minutes). The procedure was identical to the practice session, except the moderator did not provide any feedback. 

\subsection{User Study Results}
All participants were able to successfully identify the location of the single impulsive cues with over 90\% accuracy on average (97\%, 96\%, 93\% for the high, medium, and low strength cues, respectively, see Fig.~\ref{fig7}d and Fig.~S3).

We conducted a Shapiro–Wilk test on the localization accuracy for each cue strength and found that the data could not be assumed to be normally distributed. As a result, we conducted a Friedman test followed by a post hoc pairwise comparisons using the Wilcoxon signed-rank test with a significance threshold of 0.05 and Bonferroni correction for multiple comparisons. There was no statistically significant difference between the high and medium strength cues, as well as the medium and low strength cues, however, there was a statistically significant difference between the high and low strength cues ($p = 0.0156$ for a significant threshold $p = 0.0167$ after correction). Because we selected the low strength cue to be as small as possible while still being discernible based on pilot results, it is likely that some participants had a more difficult time localizing this cue due to differences in the minimum detectable threshold from user to user.

Along with the single impulse cues, participants successfully identified seven different patterns with over 90\% accuracy (see Fig.~\ref{fig7}f). Based on our study, participants struggled most when distinguishing the tap cue (v) from the heartbeat cue (vi), most likely due to the similar principle frequency of the patterns. Nevertheless, participants were able to easily identify each of the cues even with minimal training, therefore highlighting that the CASAband can be a promising tool for delivering simple and intuitive haptic guidance during real world scenarios.

\section{Haptic Guidance Demonstrations}

Recent studies have explored potential applications of haptic devices for navigation guidance, enabling users to find their way in unfamiliar locations~\cite{kappers2024hands}. Research has also suggested that haptic cues for guidance assistance has less cognitive load on the user when compared to visual and auditory notifications \cite{bouzbib2023survey}. They can also support users with impaired vision or hearing by providing clear and reliable directional cues through haptic feedback~\cite{slade2021multimodal}. 

We conducted two demonstrations highlighting potential use cases for CASAband as an information channel using haptics to assist with guidance and navigation: (1) Task guidance by conveying instructions using tactile cues while visual and auditory information was blocked; and (2) outdoor pedestrian navigation using tactile information in environments when multiple sensory stimuli coexist.

\subsection{Task Instruction Demonstration}

\begin{figure*}
  \includegraphics[width=\linewidth]{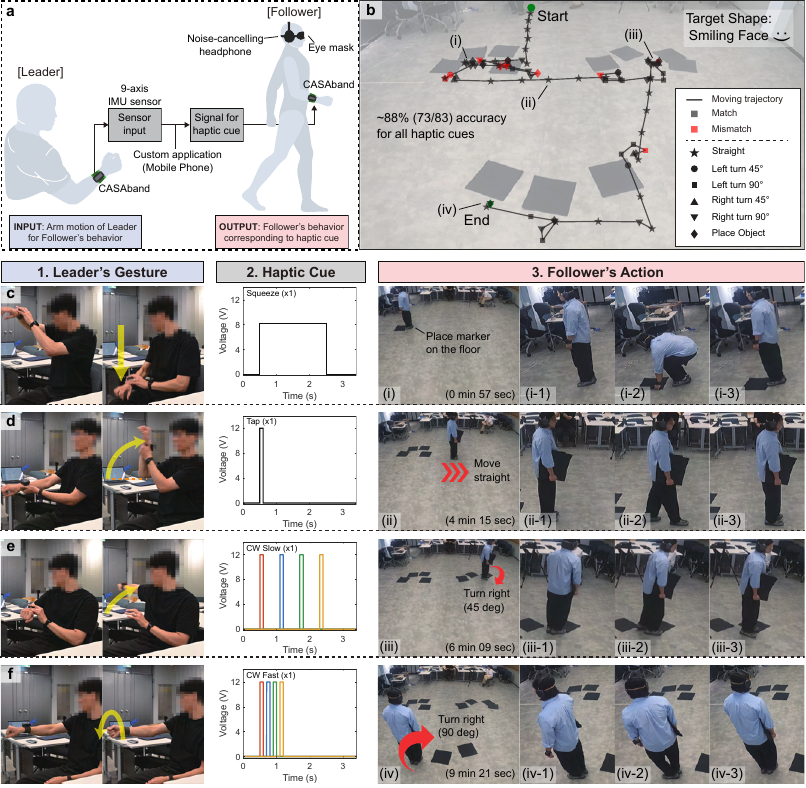}
  \vspace{-15pt}
  \caption{Pick and place task demonstration by two users equipped with a CASAband. (\textbf{a}) Schematic of the demonstration with blocked visual and auditory cues. (\textbf{b}) The trajectory that the follower walked and their accuracy when identifying the patterns delivered by the leader. The total demo time was 10 minutes. (\textbf{c}--\textbf{f}) Examples of the procedure to deliver a haptic cue to the follower. The leader completes a gesture which generates the haptic cue that is rendered to the follower. The follower then completes the action corresponding to the haptic cue. The Leader's gesture for the Follower to turn left is the reverse of the gesture used to turn right (see (e) and (f) for the right turn).}
  \label{fig8}
  \vspace{-15pt}
\end{figure*}

The goal of this demonstration was to show that two users wearing the CASAband could interact solely through gestures and haptic feedback to collaboratively draw a desired shape (i.e., a smiling face) on the floor. The demonstration involved two participants: the Leader, who directed the motion via arm gestures, and the Follower, who performed actions based on the haptic cues received (See Fig.~\ref{fig8}a). Both users wore a CASAband that communicated with a custom mobile application using Bluetooth. To limit visual and auditory information during the demonstration, the Follower wore an eye mask and noise-canceling headphones. When the Leader performed a specific arm gesture, the IMU sensor embedded in the CASAband transmitted motion data to the mobile app, which recognized the gesture and delivered the corresponding haptic pattern to the Follower.

For this demonstration, we used six pre-programmed haptic cues that were evaluated in the user study: `Tap', `Squeeze', `CW slow', `CCW slow', `CW fast', and `CCW fast'. We used the `Squeeze' cue to represent task execution (i.e., place the marker on the floor) and the other five cues for movement guidance (`CW fast': 90 degree right turn, `CW slow': 45 degree right turn, `CCW fast': 90 degree left turn, `CCW slow': 45 degree left turn, `Tap': step forward). The Follower then performed the corresponding movement or task based on each received haptic cue.

For user gesture recognition, we utilized the onboard IMU sensor (LSM9DS1) of an Arduino Nano BLE 33. The IMU data was transmitted in real-time via Bluetooth to the custom mobile application. The raw accelerometer and gyroscope signals were fused using a quaternion-based Madgwick filter to estimate the forearm orientation, which was then used to compute the total yaw change for the slow-rotation gestures. The raw sensor data was processed to classify six distinct gestures based on the following conditions: CCW Fast (X-axis gyroscope $>$ 200 $\&$ Y-axis accelerometer $<$ -0.5), CW Fast (X-axis gyroscope $<$ -200 $\&$ Y-axis accelerometer $>$ 0.5), Tapping (Z-axis accelerometer $<$ -0.5 $\&$ Y-axis gyroscope $<$ -100.0), Squeeze (Z-axis accelerometer $>$ 2.5), CCW Slow (total yaw change $>$ +50° after absolute Z-axis gyroscope $>$ 200), and CW Slow (total yaw change $<$ -50° after absolute Z-axis gyroscope $>$ 200). The coordinate system for the IMU axes was defined as in Fig.~S4. 

Prior to the demonstration, a 30-minute training session was conducted to understand the mapping between the haptic cues and the desired action. The trajectory of the Follower's movement and the locations where they placed objects show that the intended smiling face shape was successfully created, despite the absence of visual and auditory information (see Fig.~\ref{fig8}b--f). The total demonstration time was about 10 minutes, during which the Follower recognized 73 out of 83 patterns sent by the Leader (88$\%$ accuracy, see Fig.~\ref{fig8}b). Although some cues were mismatched, the real-time feedback from the Leader allowed the trajectory to be adjusted, therefore allowing the follower to still successfully complete the task.

\begin{figure*}
  \includegraphics[width=\linewidth]{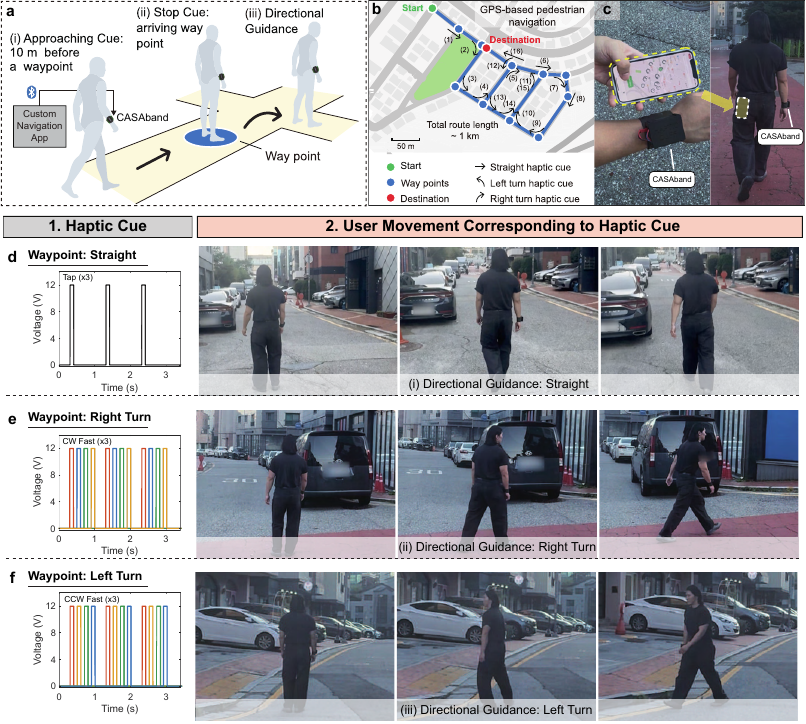}
  \vspace{-15pt}
  \caption{GPS navigation demonstration. (\textbf{a}) Schematic of the outdoor navigation demonstration using the CASAband. (\textbf{b}) The path walked by the user wearing the CASAband. The user was directed through multiple waypoints for over 1~km. (\textbf{c}) The custom application connected to the CASAband via Bluetooth for navigation guidance. In the demonstration, the user reached the destination using only haptic feedback. (\textbf{d}--\textbf{f}) Representative examples of the haptic cues for directional guidance and snapshots showing the user walking straight, turning right, and turning left.}
  \label{fig9}
  \vspace{-15pt}
\end{figure*}

\subsection{GPS Navigation Demonstration}

In the second demonstration, we conducted an outdoor navigation experiment where a user followed a predetermined route using only directional feedback from the CASAband (Fig.~\ref{fig9}a). We defined a route approximately 1 km long with a total of 16 waypoints (Fig.~\ref{fig9}b). The user’s location was tracked in real-time using the GPS information from a mobile phone, which also triggered the haptic patterns from a custom navigation application.

We developed a custom iOS application using Apple’s MapKit frameworks to obtain real-time GPS coordinates, that computed walking routes between predefined waypoints, and recorded the user’s trajectory during each trial. Bluetooth was used to wirelessly connect to the CASAband and transmit simple directional commands according to the user’s current position along the route. During experiments, the application ran on a mobile phone (iPhone 12 Pro, Apple Inc.) carried by the user, continuously logging GPS data in the background while triggering the haptic patterns as the user approached predefined decision points on the path. A more detailed view of the application layout and example user trajectory recordings can be seen in the supplementary video.

The process of providing navigation information using the CASAband was as follows. Once the demonstration began, the user placed their mobile phone in their pocket and only relied on haptic information to reach the destination (see Fig.~\ref{fig9}c). When no cue was delivered, the user continued walking straight. Upon approaching each waypoint, three types of haptic cues were sequentially provided to convey navigational instructions:\\
(i) Approaching cue: 10 meters before reaching the waypoint, the user was given advanced notice of the direction to proceed. a `Tap' haptic cue indicated straight, a `CW slow' haptic cue indicated a right turn, and a `CCW slow' haptic cue indicated a left turn.\\
(ii) Stop cue: Once within a 5-meter radius of the waypoint, a `Squeeze' haptic cue was sent to instruct the user to stop at that location.\\
(iii) Directional cue: To prevent missed or misinterpreted cues, the directional cue for the next movement was repeated three times to the stopped user. The haptic cues `Tap’, `CW fast’, and `CCW fast’ were used to inform the user to continue straight, turn right, and turn left, respectively.

The user successfully navigated the entire route and arrived at the correct destination in 14 minutes without any visual or auditory aids (see Fig.~\ref{fig6}d--f for examples of the directional cues at different waypoints). The user walked at an average speed of 4.3 km/h, which was similar to the typical walking speed of a pedestrian. This result indicated that CASAband can deliver reliable directional guidance even in outdoor environments with numerous visual and auditory distractions.

\section{Discussion}

In this work, we presented CASAband, a lightweight wearable haptic interface that delivered spatial and temporal haptic feedback on the wrist. The system integrated SMA-based actuators into a textile wristband to maintain a compact form factor suitable for daily wear while still being capable of generating a variety of spatial and temporal haptic cues across four independent channels. The overall system mass is kept under 65~g, including the onboard electronics and a battery for wireless operation, making it one of the lightest wrist- or forearm-worn haptic devices when considering all necessary components for operation (see Table \ref{tab:comparison}).

The CASA used in the device was designed by considering the force, displacement, and frequency requirements based on the application and perception experiments for wrist/forearm stimulation. The combination of SMA wire diameter, row count, and compliant beam length ratio was chosen to provide sufficient work output while maintaining a satisfactory operating frequency and form factor. Compared with our previous SMA actuator designs \cite{kim2022actuating}, which primarily focused on either force output or frequency response, this work provides an example that addresses the trade-off between mechanical performance and operating frequency necessary for wearable haptic applications.

The integration of the actuator within the multi-layered textile wristband was guided by both wearability and reliable force transmission. The outer layer of the wristband employed a higher-stiffness textile, while the inner layer in contact with the skin was more compliant to transmit force to the skin.  The width of the heat-bonded film and the overall tension when donned were also optimized to maximize the force transmitted to the skin. The textile wristband was fabricated using a simple 2D lamination strategy, therefore allowing us to rapidly manufacture the wristband with high repeatability. Given that the layers were all cut using a CO2 laser cutter, we can also use our manufacturing method to design different textile garments for other regions of the body, such as the upper arm or leg, although the performance requirements for the CASA would need to be changed accordingly.

We evaluated the CASAband for its force output, functional bandwidth, durability, and operating temperature at the point of contact with the skin. Participants could easily identify single stimuli at four locations around the wrist and distinguish between seven haptic patterns that consisted of different speeds, direction, and intensities with greater than 90$\%$ accuracy on average for all cases. This result indicated that the CASAband can transmit a meaningful amount of information within the constraints of a wrist-worn form factor.

The total weight of the CASAband was 63~g, which indicates its potential as a compact and lightweight wearable haptic system. In addition to the inherent advantage of the lightweight SMA-based actuator itself, the weight and volume of the required electronics to operate the actuators are minimal. Compared to other actuation methods used in recent wearable haptic devices, such as pneumatic and dielectric elastomer actuators, the proposed system exhibits an advantage in both weight and volume (Table \ref{tab:comparison}).

The two demonstrations we presented for the CASAband provided examples of the technical readiness of the device and how it may be used in everyday scenarios. In the first demonstration, the user received real-time task instructions through haptic patterns while visual and auditory information was blocked. With only a short training phase, the user could follow the required path and accomplish the task solely relying on the haptic feedback. In the second demonstration, the user successfully followed a 1 km route using only haptic directional guidance on the wrist, despite the presence of urban noise and visual stimuli. These examples suggest the feasibility of CASAband serving as a information channel that complements or substitutes visual and auditory information.

Although the proposed system was designed for practical use in everyday settings, there are a few remaining items that could be further improved. The durability tests confirmed basic reliability under repeated actuation, but longer-term use across a broader range of users and scenarios will likely require additional improvements in terms of actuator lifetime. In the current prototype, conventional wires are used to connect the actuators and electronics, which can increase the thickness of the textile wristband and complicate fabrication. Future designs could integrate conductive textiles \cite{matsuhisa2015printable} or stretchable electronics \cite{woodman2024stretchable} directly into the textile structure, or adopt a modular design, thereby enabling easier fabrication and maintenance.

From a control perspective, this work focused on the combination of pulse and step inputs to generate tapping, rotating, and squeezing patterns. While we generated different forces and patterns by varying the applied voltage to not limit the CASAband to binary haptic feedback, future work should introduce sensing and closed-loop control to enable measurement of the actuator’s force and displacement or skin-contact states. Our previous study combining the CASA with compact capacitive sensors demonstrated that such integration is feasible \cite{kim2023control}. Moreover, integrating additional sensors capable of recognizing user intent or gestures, such as EMG or capacitive touch sensing, would further enable its use for supporting two-way interaction scenarios.

\section{Conclusion}
In this work, we presented CASAband: a wearable haptic interface that leverages SMA actuators and a textile structure to provide spatial and temporal tactile feedback on the wrist in a compact, lightweight form factor. Through mechanical characterization, user perception experiments, and two demonstrations, we examined its potential for everyday wear and for providing tactile guidance in real-world environments. With further advances in durability, electrical integration, and additional functionality, we expect that this textile-based SMA haptic device will become a practical option for a variety of real-world applications where haptic feedback could be beneficial.

\section*{Acknowledgments}
This work was supported by the National Research Council of Science \& Technology (NST) grant by the Korea government (MSIT) (CRC23021-000), the Ministry of Trade, Industry \& Energy (MOTIE, Korea) grant by the Korea goverment (KEIT) (RS-2025-25462891), the National Science Foundation under Grant 2146095, the Natural Sciences and Engineering Research Council of Canada Graduate Research Scholarship, and the National Science Foundation Graduate Research Fellowship.

\bibliographystyle{IEEEtran}
\bibliography{MSP-template}

@article{banerjee2018soft,
  title={Soft robotics with compliance and adaptation for biomedical applications and forthcoming challenges},
  author={Banerjee, Hritwick and Tse, Zion Tsz Ho and Ren, Hongliang},
  journal={Int. J. Robot. Autom},
  volume={33},
  number={1},
  pages={68--80},
  year={2018}
}

@inproceedings{palmer2022haptic,
  title={Haptic feedback relocation from the fingertips to the wrist for two-finger manipulation in virtual reality},
  author={Palmer, Jasmin E and Sarac, Mine and Garza, Aaron A and Okamura, Allison M},
  booktitle={2022 IEEE/RSJ International Conference on Intelligent Robots and Systems (IROS)},
  pages={628--633},
  year={2022},
  organization={IEEE}
}

@article{sousa2024evaluating,
  title={Evaluating affective touch generated via a shape-memory-alloy arm-sleeve by subjective report and facial muscle activity},
  author={Sousa, Brais Gonzalez and Muthukumarana, Sachith and Rosenkranz, Robert and Altinsoy, M Ercan and Li, Shu-Chen},
  journal={IEEE Transactions on Haptics},
  volume={18},
  number={1},
  pages={188--197},
  year={2024},
  publisher={IEEE}
}

@article{moriyama2022wearable,
  title={Wearable haptic device presenting sensations of fingertips to the forearm},
  author={Moriyama, Taha and Kajimoto, Hiroyuki},
  journal={IEEE Transactions on Haptics},
  volume={15},
  number={1},
  pages={91--96},
  year={2022},
  publisher={IEEE}
}

@article{yoshida2024design,
  title={Design and evaluation of a 3-dof haptic device for directional shear cues on the forearm},
  author={Yoshida, Kyle T and Zook, Zane A and Choi, Hojung and Luo, Ming and O'Malley, Marcia K and Okamura, Allison M},
  journal={IEEE Transactions on Haptics},
  volume={17},
  number={3},
  pages={483--495},
  year={2024},
  publisher={IEEE}
}

@article{zhang2024haptic,
  title={A haptic feedback sleeve for a flight video game},
  author={Zhang, Chaozhou and Li, Min and Wu, Zonglin and Zhao, Chen-Guang and Yuan, Hua and Xie, Jun and Xu, Guanghua and Li, Jichun and Luo, Shan},
  journal={IEEE Transactions on Haptics},
  volume={18},
  number={1},
  pages={198--207},
  year={2024},
  publisher={IEEE}
}

@article{vuong2025effects,
  title={Effects of Wrist-Worn Haptic Feedback on Force Accuracy and Task Speed during a Teleoperated Robotic Surgery Task},
  author={Vuong, Brian B and Davidson, Josie and Cheon, Sangheui and Cho, Kyujin and Okamura, Allison M},
  journal={IEEE Robotics and Automation Letters},
  year={2025},
  publisher={IEEE}
}

@article{bouzbib2023survey,
  title={Survey of Wearable Haptic Technologies for Navigation Guidance},
  author={Bouzbib, Elodie and Kuang, Lisheng and Giordano, Paolo Robuffo and L{\'e}cuyer, Anatole and Pacchierotti, Claudio},
  year={2023}
}

@inproceedings{kodali2023wearable,
  title={Wearable Sensory Substitution for Proprioception via Deep Pressure},
  author={Kodali, Sreela and Vuong, Brian B and Bulea, Thomas C and Chesler, Alexander T and B{\"o}nnemann, Carsten G and Okamura, Allison M},
  booktitle={2023 IEEE World Haptics Conference (WHC)},
  pages={286--292},
  year={2023},
  organization={IEEE}
}

@article{kuchenbecker2006improving,
  title={Improving contact realism through event-based haptic feedback},
  author={Kuchenbecker, Katherine J and Fiene, Jonathan and Niemeyer, G{\"u}nter},
  journal={IEEE transactions on visualization and computer graphics},
  volume={12},
  number={2},
  pages={219--230},
  year={2006},
  publisher={IEEE}
}

@article{yu2019skin,
  title={Skin-integrated wireless haptic interfaces for virtual and augmented reality},
  author={Yu, Xinge and Xie, Zhaoqian and Yu, Yang and Lee, Jungyup and Vazquez-Guardado, Abraham and Luan, Haiwen and Ruban, Jasper and Ning, Xin and Akhtar, Aadeel and Li, Dengfeng and others},
  journal={Nature},
  volume={575},
  number={7783},
  pages={473--479},
  year={2019},
  publisher={Nature Publishing Group UK London}
}

@article{visell2009tactile,
  title={Tactile sensory substitution: Models for enaction in HCI},
  author={Visell, Yon},
  journal={Interacting with Computers},
  volume={21},
  number={1-2},
  pages={38--53},
  year={2009},
  publisher={Oxford University Press Oxford, UK}
}

@article{okamura2009haptic,
  title={Haptic feedback in robot-assisted minimally invasive surgery},
  author={Okamura, Allison M},
  journal={Current opinion in urology},
  volume={19},
  number={1},
  pages={102--107},
  year={2009},
  publisher={LWW}
}

@article{pacchierotti2023cutaneous,
  title={Cutaneous/tactile haptic feedback in robotic teleoperation: Motivation, survey, and perspectives},
  author={Pacchierotti, Claudio and Prattichizzo, Domenico},
  journal={IEEE Transactions on Robotics},
  volume={40},
  pages={978--998},
  year={2023},
  publisher={IEEE}
}

@article{jumet2023fluidically,
  title={Fluidically programmed wearable haptic textiles},
  author={Jumet, Barclay and Zook, Zane A and Yousaf, Anas and Rajappan, Anoop and Xu, Doris and Yap, Te Faye and Fino, Nathaniel and Liu, Zhen and O’Malley, Marcia K and Preston, Daniel J},
  journal={Device},
  volume={1},
  number={3},
  year={2023},
  publisher={Elsevier}
}

@article{sanchez2024cutaneous,
  title={Cutaneous Electrohydraulic (CUTE) Wearable Devices for Pleasant Broad-Bandwidth Haptic Cues},
  author={Sanchez-Tamayo, Natalia and Yoder, Zachary and Rothemund, Philipp and Ballardini, Giulia and Keplinger, Christoph and Kuchenbecker, Katherine J},
  journal={Advanced Science},
  volume={11},
  number={48},
  pages={2402461},
  year={2024},
  publisher={Wiley Online Library}
}

@article{fleck2025wearable,
  title={Wearable multi-sensory haptic devices},
  author={Fleck, Joshua J and Zook, Zane A and Clark, Janelle P and Preston, Daniel J and Lipomi, Darren J and Pacchierotti, Claudio and O’Malley, Marcia K},
  journal={Nature Reviews Bioengineering},
  pages={1--15},
  year={2025},
  publisher={Nature Publishing Group UK London}
}

@article{corniani2020tactile,
  title={Tactile innervation densities across the whole body},
  author={Corniani, Giulia and Saal, Hannes P},
  journal={Journal of Neurophysiology},
  volume={124},
  number={4},
  pages={1229--1240},
  year={2020},
  publisher={American Physiological Society Bethesda, MD}
}

@article{giraud2021haptigami,
  title={Haptigami: A fingertip haptic interface with vibrotactile and 3-DoF cutaneous force feedback},
  author={Giraud, Frederic H and Joshi, Sagar and Paik, Jamie},
  journal={IEEE Transactions on Haptics},
  volume={15},
  number={1},
  pages={131--141},
  year={2021},
  publisher={IEEE}
}

@article{ha2025full,
  title={Full freedom-of-motion actuators as advanced haptic interfaces},
  author={Ha, Kyoung-Ho and Yoo, Jaeyoung and Li, Shupeng and Mao, Yuxuan and Xu, Shengwei and Qi, Hongyuan and Wu, Hanbing and Fan, Chengye and Yuan, Hanyin and Kim, Jin-Tae and others},
  journal={Science},
  volume={387},
  number={6741},
  pages={1383--1390},
  year={2025},
  publisher={American Association for the Advancement of Science}
}

@inproceedings{choi2017grabity,
  title={Grabity: A wearable haptic interface for simulating weight and grasping in virtual reality},
  author={Choi, Inrak and Culbertson, Heather and Miller, Mark R and Olwal, Alex and Follmer, Sean},
  booktitle={Proceedings of the 30th annual ACM symposium on user interface software and technology},
  pages={119--130},
  year={2017}
}

@article{ji2021untethered,
  title={Untethered feel-through haptics using 18-$\mu$m thick dielectric elastomer actuators},
  author={Ji, Xiaobin and Liu, Xinchang and Cacucciolo, Vito and Civet, Yoan and El Haitami, Alae and Cantin, Sophie and Perriard, Yves and Shea, Herbert},
  journal={Advanced Functional Materials},
  volume={31},
  number={39},
  pages={2006639},
  year={2021},
  publisher={Wiley Online Library}
}

@article{palagi2023mechanical,
  title={A mechanical hand-tracking system with tactile feedback designed for telemanipulation},
  author={Palagi, Marcello and Santamato, Giancarlo and Chiaradia, Domenico and Gabardi, Massimiliano and Marcheschi, Simone and Solazzi, Massimiliano and Frisoli, Antonio and Leonardis, Daniele},
  journal={IEEE Transactions on Haptics},
  volume={16},
  number={4},
  pages={594--601},
  year={2023},
  publisher={IEEE}
}

@article{xiong2022so,
  title={So-EAGlove: VR haptic glove rendering softness sensation with force-tunable electrostatic adhesive brakes},
  author={Xiong, Quan and Liang, Xuanquan and Wei, Daiyue and Wang, Huacen and Zhu, Renjie and Wang, Ting and Mao, Jianjun and Wang, Hongqiang},
  journal={IEEE Transactions on Robotics},
  volume={38},
  number={6},
  pages={3450--3462},
  year={2022},
  publisher={IEEE}
}

@article{kim2015acceptance,
  title={An acceptance model for smart watches: Implications for the adoption of future wearable technology},
  author={Kim, Ki Joon and Shin, Dong-Hee},
  journal={Internet Research},
  volume={25},
  number={4},
  pages={527--541},
  year={2015},
  publisher={Emerald Group Publishing Limited}
}

@article{yin2021wearable,
  title={Wearable soft technologies for haptic sensing and feedback},
  author={Yin, Jessica and Hinchet, Ronan and Shea, Herbert and Majidi, Carmel},
  journal={Advanced Functional Materials},
  volume={31},
  number={39},
  pages={2007428},
  year={2021},
  publisher={Wiley Online Library}
}

@misc{slade2021multimodal,
  title={Multimodal sensing and intuitive steering assistance improve navigation and mobility for people with impaired vision. Science Robotics 6, 59 (2021), eabg6594},
  author={Slade, Patrick and Tambe, Arjun and Kochenderfer, Mykel J},
  year={2021}
}

@article{zhao2020wearable,
  title={A wearable soft haptic communicator based on dielectric elastomer actuators},
  author={Zhao, Huichan and Hussain, Aftab M and Israr, Ali and Vogt, Daniel M and Duduta, Mihai and Clarke, David R and Wood, Robert J},
  journal={Soft robotics},
  volume={7},
  number={4},
  pages={451--461},
  year={2020},
  publisher={Mary Ann Liebert, Inc., publishers 140 Huguenot Street, 3rd Floor New~…}
}

@article{lee2022wearable,
  title={A wearable textile-embedded dielectric elastomer actuator haptic display},
  author={Lee, Dae-Young and Jeong, Seung Hee and Cohen, Andy J and Vogt, Daniel M and Kollosche, Matthias and Lansberry, Geoffrey and Meng{\"u}{\c{c}}, Yi{\u{g}}it and Israr, Ali and Clarke, David R and Wood, Robert J},
  journal={Soft Robotics},
  volume={9},
  number={6},
  pages={1186--1197},
  year={2022},
  publisher={Mary Ann Liebert, Inc., publishers 140 Huguenot Street, 3rd Floor New~…}
}

@article{oquendo2024haptic,
  title={Haptic guidance and haptic error amplification in a virtual surgical robotic training environment},
  author={Oquendo, Yousi A and Coad, Margaret M and Wren, Sherry M and Lendvay, Thomas S and Nisky, Ilana and Jarc, Anthony M and Okamura, Allison M and Chua, Zonghe},
  journal={IEEE Transactions on Haptics},
  volume={17},
  number={3},
  pages={417--428},
  year={2024},
  publisher={IEEE}
}

@inproceedings{young2019bellowband,
  title={Bellowband: A pneumatic wristband for delivering local pressure and vibration},
  author={Young, Eric M and Memar, Amirhossein H and Agarwal, Priyanshu and Colonnese, Nick},
  booktitle={2019 IEEE World Haptics Conference (WHC)},
  pages={55--60},
  year={2019},
  organization={IEEE}
}

@inproceedings{jumet2022textile,
  title={A textile-based approach to wearable haptic devices},
  author={Jumet, Barclay and Zook, Zane A and Xu, Doris and Fino, Nathaniel and Rajappan, Anoop and Schara, Mark W and Berning, Jeffrey and Escobar, Nicolas and O'Malley, Marcia K and Preston, Daniel J},
  booktitle={2022 IEEE 5th International Conference on Soft Robotics (RoboSoft)},
  pages={741--746},
  year={2022},
  organization={IEEE}
}

@article{pezent2022design,
  title={Design, control, and psychophysics of tasbi: A force-controlled multimodal haptic bracelet},
  author={Pezent, Evan and Agarwal, Priyanshu and Hartcher-O’Brien, Jessica and Colonnese, Nicholas and O’Malley, Marcia K},
  journal={IEEE Transactions on Robotics},
  volume={38},
  number={5},
  pages={2962--2978},
  year={2022},
  publisher={IEEE}
}

@inproceedings{gupta2017hapticclench,
  title={Hapticclench: Investigating squeeze sensations using memory alloys},
  author={Gupta, Aakar and Irudayaraj, Antony Albert Raj and Balakrishnan, Ravin},
  booktitle={Proceedings of the 30th Annual ACM Symposium on User Interface Software and Technology},
  pages={109--117},
  year={2017}
}

@inproceedings{he2015pneuhaptic,
  title={PneuHaptic: delivering haptic cues with a pneumatic armband},
  author={He, Liang and Xu, Cheng and Xu, Ding and Brill, Ryan},
  booktitle={Proceedings of the 2015 ACM international symposium on wearable computers},
  pages={47--48},
  year={2015}
}

@inproceedings{culbertson2018social,
  title={A social haptic device to create continuous lateral motion using sequential normal indentation},
  author={Culbertson, Heather and Nunez, Cara M and Israr, Ali and Lau, Frances and Abnousi, Freddy and Okamura, Allison M},
  booktitle={2018 IEEE haptics symposium (HAPTICS)},
  pages={32--39},
  year={2018},
  organization={IEEE}
}

@inproceedings{hamdan2019springlets,
  title={Springlets: Expressive, flexible and silent on-skin tactile interfaces},
  author={Hamdan, Nur Al-huda and Wagner, Adrian and Voelker, Simon and Steimle, J{\"u}rgen and Borchers, Jan},
  booktitle={Proceedings of the 2019 CHI conference on human factors in computing systems},
  pages={1--14},
  year={2019}
}

@inproceedings{zhu2020pneusleeve,
  title={Pneusleeve: In-fabric multimodal actuation and sensing in a soft, compact, and expressive haptic sleeve},
  author={Zhu, Mengjia and Memar, Amirhossein H and Gupta, Aakar and Samad, Majed and Agarwal, Priyanshu and Visell, Yon and Keller, Sean J and Colonnese, Nicholas},
  booktitle={Proceedings of the 2020 CHI conference on human factors in computing systems},
  pages={1--12},
  year={2020}
}

@article{meli2018hbracelet,
  title={The hBracelet: A wearable haptic device for the distributed mechanotactile stimulation of the upper limb},
  author={Meli, Leonardo and Hussain, Irfan and Aurilio, Mirko and Malvezzi, Monica and O’Malley, Marcia K and Prattichizzo, Domenico},
  journal={IEEE Robotics and Automation Letters},
  volume={3},
  number={3},
  pages={2198--2205},
  year={2018},
  publisher={IEEE}
}

@article{sepehri2025bundled,
  title={Bundled Liquid Crystal Elastomer Actuators With Integrated Cooling for Mesoscale Soft Robots},
  author={Sepehri, Anoush and Kim, Sukjun and Agrawal, Devyansh and Yared, Hannah and Dong, Gaoweiang and Cai, Shengqiang and Morimoto, Tania K},
  journal={IEEE Robotics and Automation Letters},
  year={2025},
  publisher={IEEE}
}

@article{du2024haptiknit,
  title={Haptiknit: Distributed stiffness knitting for wearable haptics},
  author={du Pasquier, Cosima and Tessmer, Lavender and Scholl, Ian and Tilton, Liana and Chen, Tian and Tibbits, Skylar and Okamura, Allison},
  journal={Science Robotics},
  volume={9},
  number={97},
  pages={eado3887},
  year={2024},
  publisher={American Association for the Advancement of Science}
}

@article{oh2023easy,
  title={Easy-to-wear auxetic SMA knot-architecture for spatiotemporal and multimodal haptic feedbacks},
  author={Oh, Saewoong and Song, Tae-Eun and Mahato, Manmatha and Kim, Ji-Seok and Yoo, Hyunjoon and Lee, Myung-Joon and Khan, Mannan and Yeo, Woon-Hong and Oh, Il-Kwon},
  journal={Advanced Materials},
  volume={35},
  number={47},
  pages={2304442},
  year={2023},
  publisher={Wiley Online Library}
}

@inproceedings{forman2023fiberobo,
  title={FibeRobo: Fabricating 4D fiber interfaces by continuous drawing of temperature tunable liquid crystal elastomers},
  author={Forman, Jack and Kilic Afsar, Ozgun and Nicita, Sarah and Lin, Rosalie Hsin-Ju and Yang, Liu and Hofmann, Megan and Kothakonda, Akshay and Gordon, Zachary and Honnet, Cedric and Dorsey, Kristen and others},
  booktitle={Proceedings of the 36th Annual ACM Symposium on User Interface Software and Technology},
  pages={1--17},
  year={2023}
}

@inproceedings{biggs2002tangential,
  title={Tangential versus normal displacements of skin: Relative effectiveness for producing tactile sensations},
  author={Biggs, James and Srinivasan, Mandayam A},
  booktitle={Proceedings 10th Symposium on Haptic Interfaces for Virtual Environment and Teleoperator Systems. HAPTICS 2002},
  pages={121--128},
  year={2002},
  organization={IEEE}
}

@inproceedings{erwin2014design,
  title={Design and perceptibility of a wearable haptic device using low-frequency stimulations on the forearm},
  author={Erwin, Andrew and Sup, Frank},
  booktitle={2014 IEEE Haptics Symposium (HAPTICS)},
  pages={505--508},
  year={2014},
  organization={IEEE}
}

@article{nolan1982two,
  title={Two-point discrimination assessment in the upper limb in young adult men and women},
  author={Nolan, Michael F},
  journal={Physical therapy},
  volume={62},
  number={7},
  pages={965--969},
  year={1982},
  publisher={Oxford University Press}
}

@article{kim2022actuating,
  title={Actuating compact wearable augmented reality devices by multifunctional artificial muscle},
  author={Kim, Dongjin and Kim, Baekgyeom and Shin, Bongsu and Shin, Dongwook and Lee, Chang-Kun and Chung, Jae-Seung and Seo, Juwon and Kim, Yun-Tae and Sung, Geeyoung and Seo, Wontaek and others},
  journal={Nature communications},
  volume={13},
  number={1},
  pages={4155},
  year={2022},
  publisher={Nature Publishing Group UK London}
}

@article{zhang2019robotic,
  title={Robotic artificial muscles: Current progress and future perspectives},
  author={Zhang, Jun and Sheng, Jun and O’Neill, Ciar{\'a}n T and Walsh, Conor J and Wood, Robert J and Ryu, Jee-Hwan and Desai, Jaydev P and Yip, Michael C},
  journal={IEEE transactions on robotics},
  volume={35},
  number={3},
  pages={761--781},
  year={2019},
  publisher={IEEE}
}

@article{sparks2015use,
  title={Use of silicone materials to simulate tissue biomechanics as related to deep tissue injury},
  author={Sparks, Jessica L and Vavalle, Nicholas A and Kasting, Krysten E and Long, Benjamin and Tanaka, Martin L and Sanger, Phillip A and Schnell, Karen and Conner-Kerr, Teresa A},
  journal={Advances in skin \& wound care},
  volume={28},
  number={2},
  pages={59--68},
  year={2015},
  publisher={LWW}
}

@article{liu2020actuation,
  title={Actuation frequency modeling and prediction for shape memory alloy actuators},
  author={Liu, Xiaolong and Liu, Hui and Tan, Jindong},
  journal={IEEE/ASME Transactions on Mechatronics},
  volume={26},
  number={3},
  pages={1536--1546},
  year={2020},
  publisher={IEEE}
}

@article{tadesse2010tailoring,
  title={Tailoring the response time of shape memory alloy wires through active cooling and pre-stress},
  author={Tadesse, Yonas and Thayer, Nicholas and Priya, Shashank},
  journal={Journal of Intelligent Material Systems and Structures},
  volume={21},
  number={1},
  pages={19--40},
  year={2010},
  publisher={Sage Publications Sage UK: London, England}
}

@article{ribot1996alteration,
  title={Alteration of human cutaneous afferent discharges as the result of long-lasting vibration},
  author={Ribot-Ciscar, E and Roll, JP and Tardy-Gervet, MF and Harlay, F},
  journal={Journal of Applied Physiology},
  volume={80},
  number={5},
  pages={1708--1715},
  year={1996}
}

@article{sofia2013mechanical,
  title={Mechanical and psychophysical studies of surface wave propagation during vibrotactile stimulation},
  author={Sofia, Katherine O and Jones, Lynette},
  journal={IEEE transactions on haptics},
  volume={6},
  number={3},
  pages={320--329},
  year={2013},
  publisher={IEEE}
}

@article{culbertson2018haptics,
  title={Haptics: The present and future of artificial touch sensation},
  author={Culbertson, Heather and Schorr, Samuel B and Okamura, Allison M},
  journal={Annual review of control, robotics, and autonomous systems},
  volume={1},
  number={1},
  pages={385--409},
  year={2018},
  publisher={Annual Reviews}
}

@article{woodman2024stretchable,
  title={Stretchable Arduinos embedded in soft robots},
  author={Woodman, Stephanie J and Shah, Dylan S and Landesberg, Melanie and Agrawala, Anjali and Kramer-Bottiglio, Rebecca},
  journal={Science Robotics},
  volume={9},
  number={94},
  pages={eadn6844},
  year={2024},
  publisher={American Association for the Advancement of Science}
}

@article{saal2014touch,
  title={Touch is a team effort: interplay of submodalities in cutaneous sensibility},
  author={Saal, Hannes P and Bensmaia, Sliman J},
  journal={Trends in neurosciences},
  volume={37},
  number={12},
  pages={689--697},
  year={2014},
  publisher={Elsevier}
}

@inproceedings{ploch2017comparing,
  title={Comparing haptic and audio navigation cues on the road for distracted drivers with a skin stretch steering wheel},
  author={Ploch, Christopher J and Bae, Jung Hwa and Ploch, Caitlin C and Ju, Wendy and Cutkosky, Mark R},
  booktitle={2017 IEEE World Haptics Conference (WHC)},
  pages={448--453},
  year={2017},
  organization={IEEE}
}

@inproceedings{van2004waypoint,
  title={Waypoint navigation on land: Different ways of coding distance to the next waypoint},
  author={Van Veen, HAHC and Spap{\'e}, Migh and Van Erp, JBF},
  booktitle={Proceedings of EuroHaptics},
  volume={2004},
  pages={160--165},
  year={2004}
}

@article{yao2007experimental,
  title={Experimental study on skin temperature and thermal comfort of the human body in a recumbent posture under uniform thermal environments},
  author={Yao, Ye and Lian, Zhiwei and Liu, Weiwei and Shen, Qi},
  journal={Indoor and Built Environment},
  volume={16},
  number={6},
  pages={505--518},
  year={2007},
  publisher={Sage Publications Sage UK: London, England}
}

@article{kappers2024hands,
  title={Hands-free haptic navigation devices for actual walking},
  author={Kappers, Astrid ML and Holt, Raymond J and Junggeburth, Tessa JW and Oen, Max Fa Si and van de Wetering, Bart JT and Plaisier, Myrthe A},
  journal={IEEE Transactions on Haptics},
  volume={17},
  number={4},
  pages={528--545},
  year={2024},
  publisher={IEEE}
}

@article{matsuhisa2015printable,
  title={Printable elastic conductors with a high conductivity for electronic textile applications},
  author={Matsuhisa, Naoji and Kaltenbrunner, Martin and Yokota, Tomoyuki and Jinno, Hiroaki and Kuribara, Kazunori and Sekitani, Tsuyoshi and Someya, Takao},
  journal={Nature communications},
  volume={6},
  number={1},
  pages={7461},
  year={2015},
  publisher={Nature Publishing Group UK London}
}

@inproceedings{sepehri2025retrofitting,
  title={Retrofitting Soft Assistive Robots with Sew-Free Sensing Garments for Joint Motion Tracking and Kinematic Feedback},
  author={Sepehri, Anoush and Tolley, Michael T and Morimoto, Tania K},
  booktitle={2025 International Conference On Rehabilitation Robotics (ICORR)},
  pages={1--7},
  year={2025},
  organization={IEEE}
}

@inproceedings{kim2023control,
  title={Control of Shape Memory Alloy Actuator via Electrostatic Capacitive Sensor for Meso-scale Mirror Tilting System},
  author={Kim, Baekgyeom and Lee, Doohoe and Kim, Dongjin and Han, Seungyong and Kang, Daeshik and Kim, Uikyum and Koh, Je-sung},
  booktitle={2023 IEEE International Conference on Robotics and Automation (ICRA)},
  pages={2634--2640},
  year={2023},
  organization={IEEE}
}

\end{document}